\documentclass[usenatbib]{mn2e}

\usepackage{graphicx}
\usepackage{epsfig}

\def\Lya{Ly$\alpha$~}
\def\OM{$\Omega_{\rm m}$~} 

\def\OB{$\Omega_{\rm b}$~} 
\def\LCDM{$\Lambda$CDM~} 
\def\gam{$\Gamma_{\rm HI}$~}
\def\OBh{$\Omega_{\rm b}h^{2}$~}
\def\gamn{$\Gamma_{\rm -12}$~}
\def\sig{$\sigma_{\rm 8}$~} 
\def\Fobs{$\langle F \rangle_{\rm obs}$~}

\def\teff{$\tau_{\rm eff}$~}

\def\HeII{\hbox{He~$\rm \scriptstyle II\ $}~}

\def\MgII{\hbox{Mg~$\rm \scriptstyle II\ $}~}
\def\OIII{\hbox{[O~$\rm \scriptstyle III$}]~}
\def\SiIV{\hbox{Si~$\rm \scriptstyle IV$}~}
\def\CIV{\hbox{C~$\rm \scriptstyle IV$}~}

\title[The Lyman-$\alpha$ forest opacity and the metagalactic hydrogen ionization rate]{The Lyman-$\alpha$ forest opacity and the metagalactic
       hydrogen ionization rate at $z\sim 2-4$}

\author[J.S. Bolton, M.G. Haehnelt, M. Viel \& V. Springel]
{James S. Bolton$^{1}$\thanks{E-mail:jsb@ast.cam.ac.uk}, Martin G. Haehnelt$^{1,2}$, Matteo Viel$^{1}$ and Volker Springel$^{3}$\\
  $^{1}$Institute of Astronomy, University of Cambridge, Madingely
  Road, Cambridge, CB3 0HA\\
  $^{2}$ Kavli Institute for Theoretical Physics,
  University of California, Santa Barbara, CA 93106, USA\\
  $^{3}$ Max-Planck-Institut f\"{u}r Astrophysik,
  Karl-Schwarzschild-Str. 1, Garching bei M\"{u}nchen, Germany\\}

\begin{document}

\date{25 November 2004}

\maketitle

\label{firstpage}

\begin{abstract}
  Estimates of the metagalactic hydrogen ionization rate from the \Lya
  forest opacity in QSO absorption spectra depend on the complex
  density distribution of neutral hydrogen along the line-of-sight.
  We use a large suite of high resolution hydrodynamical simulations
  to investigate in detail the dependence of such estimates on
  physical and numerical parameters in the context of \LCDM models.
  Adopting fiducial values for cosmological parameters together with
  published values of the temperature of the IGM and the effective
  optical depth, the metagalactic ionization rates which reproduce the
  \Lya effective optical depth at $z=[2,3,4]$ are $\Gamma_{\rm HI} =
  [1.3\pm^{0.8}_{0.5}, 0.9\pm0.3, 1.0\pm^{0.5}_{0.3} ] \times
  10^{-12}$ $\rm s^{-1}$, respectively.  The errors include estimates
  of uncertainties in the relevant physical parameters and the
  numerical accuracy of the simulations.  We find the errors are
  dominated by the uncertainty in the temperature of the low-density
  IGM.  The estimated metagalactic hydrogen ionization rate for the
  neutral hydrogen distribution in the current concordance \LCDM model
  is more than four times the value inferred for that in an
  Einstein-de Sitter model of the same r.m.s. density fluctuation
  amplitude $\sigma_{8}$. The estimated ionization rate is also more
  than double that expected from updated estimates of the emissivity
  of observed QSOs alone. A substantial contribution from galaxies
  appears to be required at all redshifts.
\end{abstract}
 

\section{Introduction}\label{sec:intro}

In recent years, high resolution, high signal-to-noise quasi-stellar
object (QSO) spectra ({\it e.g.} Lu et al. 1996; Kirkman \& Tytler
1997; Kim et al. 2002), combined with hydrodynamical simulations
\citep{b13,b57,b56,b58,b12}, have provided fresh insight into the
nature of the intergalactic medium (IGM).  It is now widely accepted
that the \Lya forest arises from absorption by density fluctuations
imposed on the warm photoionized IGM as suggested by analytical
calculations (Bi, B\"{o}rner \& Chu 1992, Bi \& Davidson 1997), and is
a natural consequence of hierarchical structure formation within
CDM-like models (see Rauch 1998 for a review).

One notable inconsistency with this paradigm has been the rather low
values of \gam inferred from early hydrodynamical simulations of the
\Lya forest if scaled to appropriate assumptions for the IGM
temperature ({\it e.g.} Rauch et al. 1997).  However, the matter
density $\Omega_{\rm m}$, baryon density $\Omega_{\rm b}$, amplitude
of mass fluctuations \sig and the parameterised Hubble constant
$h=H_{0}/100$ $\rm kms^{-1}Mpc^{-1}$, as well as the gas temperature
and the \Lya effective optical depth, \teff$= -$ln$\langle F
\rangle_{\rm obs}$, all influence such an estimate.  This has led to a
wide range of estimates for \gam inferred from the \Lya forest opacity
using numerical simulations with different assumptions (Rauch et al.
1997; Theuns et al. 1998; Haehnelt et al. 2001; McDonald \& Miralda
Escud\'{e} 2001; Hui et al.  2002; Schaye et al. 2003; Sokasian, Abel
\& Hernquist 2003; Meiksin \& White 2003, 2004; Tytler et al. 2004).
This complicates the comparison with determinations of the ionization
rate from estimates of the integrated emission from observed QSOs
and/or galaxies ({\it e.g.}  Haardt \& Madau 1996, hereafter HM96;
Madau, Haardt \& Rees 1999) and estimates using the proximity effect
({\it e.g.} Bajtlik, Duncan \& Ostriker 1988, Bechtold 1994, Cooke,
Espey \& Carswell 1997, Scott et al. 2000) which both also have rather
large uncertainties.  Furthermore, \citet{b23}, in a study of Lyman
break galaxies at $z \simeq 3.4$, suggested that the intensity of the
ionizing background may be a factor of a few larger than in the model
of HM96 due to a large contribution from star-forming galaxies.
 
In this paper we present new estimates of the metagalactic hydrogen
ionization rate using \Lya forest opacity measurements taken from the
literature.  Our approach expands on previous studies in that we
examine the dependence of \gam upon the various cosmological and
astrophysical parameters using a large suite of hydrodynamical
simulations.  In section~\ref{sec:code} we present an overview of our
numerical code and our method for obtaining \gam from artificial \Lya
spectra.  We examine how \gam depends on a wide variety of parameters
in section~\ref{sec:scale}, and compare our results to previous
estimates using the same technique in section~\ref{sec:comp}.  In
section~\ref{sec:results} we present our best values for \gamn at
$z=[2,3,4]$ and consider the implications for sources of ionizing
radiation.  Section~\ref{sec:cons} contains our summary and
conclusions.

\section{From mean flux to ionization rate} \label{sec:code}
\subsection{The fluctuating Gunn-Peterson approximation}

At redshifts where the \Lya forest is commonly observed, density
fluctuations in the low density, unshocked baryons closely follow
those of the underlying dark matter \citep{b10}.  The balance between
photoheating of unshocked gas and cooling due to the adiabatic
expansion of the Universe results in a power law temperature-density
relation for the normalised gas density $\Delta = \rho/\langle \rho
\rangle \leq 10$, where $T = T_{0} \Delta^{\gamma -1}$ \citep{b2}.
This is often referred to as the effective equation of state of the
IGM, for which typically, $ 10^{3.7}$ $\rm K$ $\leq T_{0} \leq 10^{4.3}$ $\rm
K$ and $ 1.0 \leq \gamma \leq 1.6$.  Additionally, in photoionization
equilibrium the \Lya optical depth, $\tau$, is proportional to the
density of neutral hydrogen, $n_{\rm HI}$.  Neglecting collisional
ionization, which is an acceptable approximation at the low densities
associated with \Lya forest absorbers, $ n_{\rm HI} \propto
\rho^{2}T^{-0.7}/\Gamma_{\rm HI} $, where $\rho$ is the gas density,
$T$ is the temperature and $\Gamma_{\rm HI}$ is the hydrogen
ionization rate per atom.  Combining these relations gives a simple
power law expression relating the optical depth of \Lya absorbers to
the physical properties of the hydrogen gas \citep{b6}.  This relation
has been termed the Fluctuating Gunn-Peterson Approximation (FGPA).

The optical depth due to \Lya absorption for a homogeneous
distribution of hydrogen gas in an isotropically expanding Universe
was discussed by \citet{b17}.  The FGPA applies to an inhomogeneous
IGM in photoionization equilibrium.  Ignoring the effect of peculiar
velocities and thermal broadening, the optical depth to \Lya
scattering is given by ({\it e.g.} Weinberg et al. 1999, McDonald \&
Miralda-Escud\'{e} 2001):

\begin{equation} \tau = \tau_{0} \frac{(1 + z)^{6} (\Omega_{b} h^{2})^{2}}{T^{0.7} H(z) \Gamma_{-12}(z)} \Delta^{\beta} \label{eq:FGPA}  \end{equation}

\noindent 
where \OB is the baryonic matter density as a fraction of the critical
density, $ h = H_{0}/100$ $\rm kms^{-1}Mpc^{-1} $ for the present day
Hubble constant $H_{0}$, $T$ is the gas temperature at normalised
density $\Delta$, $\Gamma_{-12}(z) = \Gamma_{\rm HI}(z)/10^{-12}$ $\rm
s^{-1}$ at redshift $z$ and $\tau_{0}$ is a constant.  The index
$\beta = 2 - 0.7(\gamma - 1)$ is determined by the slope of the
effective equation of state, and asymptotically approaches $\simeq
1.6$ after reionization.  For a flat Universe at redshifts $z \geq 2$
the Hubble parameter can be approximated by $H(z) \simeq
H_{0}\Omega_{\rm m}^{1/2}(1 + z)^{3/2}$.  Thus, given the correct
density distribution, velocity distribution and effective equation of
state at redshift $z$, the \Lya optical depth should scale according
to the parameter combination:

\begin{equation} \mu =  \Omega_{\rm b}^{2}h^{3}T^{-0.7}\Omega_{\rm
m}^{-0.5} \Gamma_{\rm -12}^{-1} \label{eq:mu}  
\end{equation}

\noindent
for redshifts $z \geq 2$ (Rauch et al. 1997).  This assumes that
thermal broadening and peculiar velocities will not have a large
effect on the optical depth distribution. At the densities and
physical scales of the \Lya forest absorbers Hubble broadening is
expected to be dominant \citep{b52} which should make this a
reasonable assumption.

We match the mean flux of artificial \Lya spectra constructed from
hydrodynamical simulations to observed values by linearly rescaling
the simulated optical depths. If the FGPA approximation holds this
corresponds to changing the parameter combination $\mu$ by the same
factor.  We have determined $\mu$ in this way.  If independent
estimates are available for $\Omega_{\rm b}$, $\Omega_{\rm m}$, $h$,
$T$ and \teff, the magnitude of \gamn required to reproduce the
observed mean flux can be derived.  This assumes that the effect of
absorption by fully neutral gas on the mean flux level is small, which
should again be a reasonable assumption.

\subsection{Numerical code}

We proceed to investigate how \gamn depends on several cosmological
and astrophysical parameters using a suite of hydrodynamical
simulations.  The simulations were run using a new version of the
parallel TreeSPH code GADGET \citep{b25}.  GADGET-2 has the option of
using a TreePM mode to perform the long range gravitational
computations, which results in a significant reduction in
computational run-time.

The conservative entropy formulation of \citet{b26} is used to model
the SPH particles, and radiative cooling is included using the
prescription of \citet{b11}.  We have used the UV background specified
by HM96 based on observed counts of QSOs, but with artificially
increased heating rates to explore the effect of gas temperature on
$\Gamma_{-12}$.  Our fiducial model uses a heating rate a factor of
$3.3$ greater than the HM96 value in the optically thin limit which
results in temperatures close to those observed in the IGM
\citep{b27,b28,b29}.  The fiducial temperatures at mean gas density we
shall assume are $T_{0}=[11200,17800,12500]$ $\rm K$ at $z=[2,3,4]$,
based on the results of \citet{b27} which are consistent with \HeII
reionization at $z\simeq3$.  At present there are only weak
observational constraints on the index $\gamma$ of the temperature
density relation and its dependence on redshift \citep{b29,b27,b48}.
The plausible range is $1<\gamma<1.6$ and in the following we will
adopt $\gamma= 1.3$ as the fiducial value.  Unless otherwise noted the
temperatures of the gas have been rescaled such that the temperature
at mean gas density $T_{0}$ has the above fiducial values and the
index of the power-law temperature-density relation for low gas
densities is $\gamma= 1.3$.  At higher gas densities the
temperature-density relation asymptotes to $T\simeq10^{4}$ $\rm K$ as
atomic cooling becomes important.
 
Star formation is included using a simplified prescription to maximise
the speed of the simulations.  All gas particles with $\Delta >
10^{3}$ and $T < 10^{5}$ $\rm K$ are converted into collisionless
stars.  Comparisons with identical simulations using the multi-phase
model of \citet{b32} show little difference in the \Lya absorption
\citep{b8}.  Additionally, all GADGET-2 feedback options have been
disabled.  The effect of feedback from galactic winds on the flux
distribution is uncertain but is believed to be small ({\it e.g.}
Theuns et al. 2002; Bruscoli et al. 2003; Desjacques et al. 2004).
Simulations run by one of us (VS) indicate that the inclusion of winds
only marginally alters the opacity of the \Lya forest, such that the
mean transmitted \Lya flux predicted by different models varies to
less than half a per cent.  There also appears to be no monotonic
trend of the mean transmitted flux with wind strength.  It is,
however, conceivable that more realistic implementations of galactic
winds within our simulations may lead to a multi-phase medium with
dense cool clouds, which may increase the \Lya forest opacity and thus
the estimated hydrogen ionization rate.  We neglect this possibility
in this work.  Note that Theuns et al. (2002) find the inclusion of
winds alters the mean transmission by $10$ per cent.

The simulation parameters are listed in Table~\ref{tab:param}.  For
each model we have run a simulation with $200^{3}$ dark matter
particles and $200^{3}$ gas particles within a $15h^{-1}$ comoving Mpc
box with periodic boundary conditions.  We find this to be the best
compromise between accuracy and speed for our parameter study.  For
our fiducial model we have also run simulations with different box
sizes and mass resolutions to test numerical convergence (see
section~\ref{sec:res}).  These are listed in Table~\ref{tab:param2}.
Note that the simulations with $200^{3}$ particles used for the
parameter study do not give fully converged estimates of \gamn and we
will later show results corrected for box size and resolution.  In
addition, we have run \LCDM and SCDM simulations with $2 \times
64^{3}$ particles and box sizes of $10h^{-1}$ and $11.1h^{-1}$
comoving Mpc respectively, to mimic the simulations used in the
seminal study of Rauch et al.  (1997). The parameters are also listed
in Table~\ref{tab:param}.

\begin{table} 
\centering
\caption{
 Parameters used in our suite of hydrodynamical simulations where we
assume a flat Universe with $\Omega_{\Lambda} = 1- \Omega_{\rm m}$.  The
last column lists the photoheating rate as a factor of the HM96 value.}

\begin{tabular}{c|c|c|c|c|c|c|c}
  \hline
    Name & \OM & \OBh & $h$ & \sig & $n$ & UV \\  
  \hline
 Fiducial      & 0.26 & 0.0240 & 0.72 & 0.85 & 0.95 & 3.3 \\
 T1            & 0.26 & 0.0240 & 0.72 & 0.85 & 0.95 & 0.3 \\ 
 T2            & 0.26 & 0.0240 & 0.72 & 0.85 & 0.95 & 1.0 \\ 
 T3            & 0.26 & 0.0240 & 0.72 & 0.85 & 0.95 & 2.0 \\
 T4            & 0.26 & 0.0240 & 0.72 & 0.85 & 0.95 & 4.0 \\
 M1            & 0.17 & 0.0240 & 0.72 & 0.85 & 0.95 & 3.3 \\
 M2            & 0.40 & 0.0240 & 0.72 & 0.85 & 0.95 & 3.3 \\
 M3            & 0.70 & 0.0240 & 0.72 & 0.85 & 0.95 & 3.3 \\
 M4            & 1.00 & 0.0240 & 0.72 & 0.85 & 0.95 & 3.3 \\
 S1            & 0.26 & 0.0240 & 0.72 & 0.50 & 0.95 & 3.3 \\
 S2            & 0.26 & 0.0240 & 0.72 & 0.70 & 0.95 & 3.3 \\
 S3            & 0.26 & 0.0240 & 0.72 & 1.00 & 0.95 & 3.3 \\
 S4            & 0.26 & 0.0240 & 0.72 & 1.20 & 0.95 & 3.3 \\
 \hline
 $\Lambda$CDM  & 0.40 & 0.0150 & 0.65 & 0.79 & 0.95 & 3.3 \\
 SCDM          & 1.00 & 0.0125 & 0.50 & 0.70 & 1.00 & 1.0 \\
 \hline
 \label{tab:param}
\end{tabular}
\end{table}

\begin{table} 
\centering
  \caption{Resolution and box size of our six additional simulations
  which have the same parameters as the  15-200 model used for
  our parameter  study.  The mass resolution used for our parameter study is also listed for comparison.}
  \begin{tabular}{c|c|c|c|c|c|c|c}
    \hline
    Name     & Box size       & SPH         & Mass resolution \\  
             & comoving Mpc   &  particles  & $h^{-1}M_{\odot}$/gas particle \\
  \hline
  15-400     & 15$h^{-1}$     & $400^{3}$   & $6.78 \times 10^{5}$ \\
  30-400     & 30$h^{-1}$     & $400^{3}$   & $5.42 \times 10^{6}$ \\
  15-100     & 15$h^{-1}$     & $100^{3}$   & $4.34 \times 10^{7}$ \\
  30-200     & 30$h^{-1}$     & $200^{3}$   & $4.34 \times 10^{7}$ \\
  60-400     & 60$h^{-1}$     & $400^{3}$   & $4.34 \times 10^{7}$ \\
  30-100     & 30$h^{-1}$     & $100^{3}$   & $3.47 \times 10^{8}$ \\
  \hline
  15-200        & 15$h^{-1}$     & $200^{3}$   & $5.42 \times 10^{6}$ \\
  \hline
  
  \label{tab:param2}
\end{tabular}
\end{table}

The simulations were started at $z=99$, with initial conditions
generated using the CDM transfer functions of \citet{b30}.  The
initial gas temperature was $T=227$ $\rm K$ and the gravitational
softening length was $2.5h^{-1}$ comoving kpc.  The simulations were
run on COSMOS, a 152 Gb shared memory Altix 3700 with 152 1.3 GHz
Itanium2 processors hosted at the Department of Applied Mathematics
and Theoretical Physics (Cambridge).

\begin{figure*}
  \centering
\begin{minipage}{180mm}

  \psfig{figure=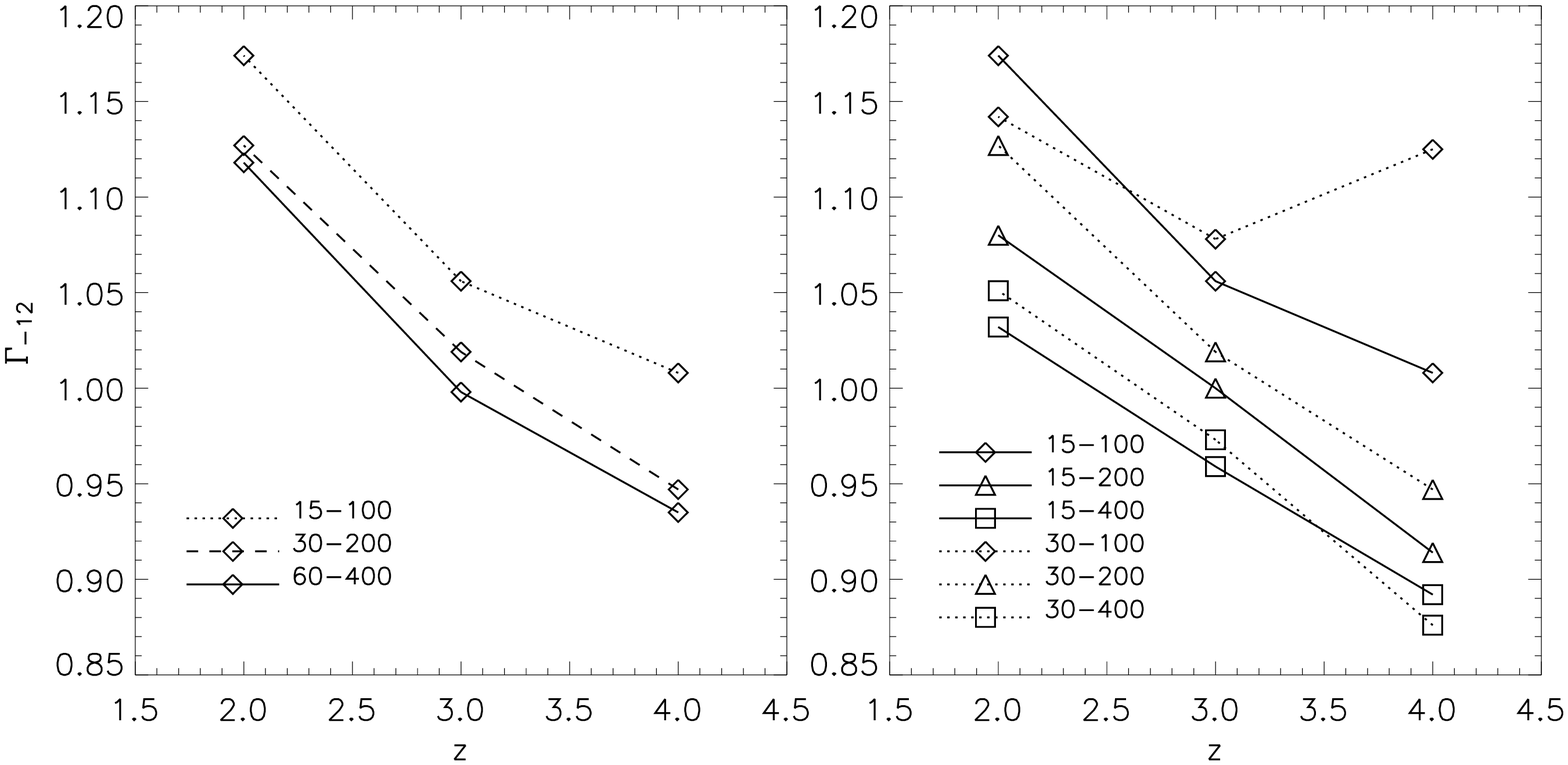,width=1.0\textwidth}
\caption{
  The effect of box size (left panel) and mass resolution (right
  panel) on the estimated $\Gamma_{-12}$ (see table~\ref{tab:param2}).
  The model parameters are those of the fiducial model (see
  table~\ref{tab:param}).}
\label{fig:res}
\end{minipage}
\end{figure*}

\subsection{Method for determining $\Gamma_{-12}$ from simulations}

Artificial \Lya spectra are constructed at $z=[2,3,4]$ from $1024$
random lines-of-sight (LOS) parallel to the box boundaries ({\it e.g.}
Theuns et al. 1998).  Each LOS consists of $1024$ pixels.  We mimic
the noise on high resolution QSO spectra by adding Gaussian
distributed noise with a total $S/N=30$.  The simulated optical depths
we use are divided by a constant factor of 1.225 to correct the
hydrogen recombination coefficient $\alpha_{\rm HI}$ used in our
simulations, which is taken from \citet{b13}.  As noted by Rauch et
al., this value of $\alpha_{\rm HI}$ is around $20\%$ larger than the
calculation of \citet{b31} within the relevant temperature range for
unsaturated absorption in the \Lya forest.  Abel et al. base their fit
for $\alpha_{\rm HI}$ on the data of Ferland et al. (1992).

To determine \gamn we rescale the optical depths of the artificial
spectra to match the mean observed flux of the \Lya forest at
$z=[2,3,4]$.  We use the central values of \teff from the fitting
formula of \citet{b41}, based on a sample of $19$ high resolution QSO
spectra, such that: $\langle F \rangle_{\rm obs} = e^{-\tau_{\rm eff}}
= [0.878\pm0.019,0.696\pm0.025,0.447\pm0.031]$.  The uncertainties on
these values were estimated by binning the $1\sigma$ errors on \teff
presented in Table 5 of \citet{b41} into redshift bins of width
$\Delta z=0.5$.  These values have been corrected for absorption from
metal lines and \Lya systems with damping wings.  We rescale the
optical depth in each pixel by a constant factor $A$ to reproduce
\Fobs at the appropriate redshift.  Hence, for a set of spectra with a
total of $N$ pixels:

\begin{equation} \langle F \rangle_{\rm obs} = \frac{1}{N} \sum_{j=1}^{N}  e^{-A\tau_{\rm j}} = e^{-\tau_{\rm eff}} \label{eq:Fobs} \end{equation}

\noindent
where $\tau_{\rm j}$ is optical depth in each pixel of the simulated
spectra.  The value of \gamn required to reproduce \Fobs from the
simulated spectra is then calculated.

\subsection{The effect of box size and mass resolution on the ionization rate} \label{sec:res}

We firstly vary the box size and resolution of simulations of our
fiducial model to assess their effect on the value of \gamn.  The
simulation box must be large enough to include long wavelength density
perturbations and represent a fair sample of the Universe while
retaining adequate resolution.  The left panel of fig.~\ref{fig:res}
shows \gamn for different box sizes at fixed mass resolution.  As the
box size increases, the simulated optical depths decrease; low density
gas responsible for \Lya absorption falls into the haloes formed in
the simulation, producing larger and emptier voids.  A smaller \gamn
is thus required to reproduce the observed optical depth. The data
have almost converged for a box size of $30h^{-1}$ comoving Mpc, with
a difference of less than $2$ per cent between the \gamn inferred from
the $60-400$ and $30-200$ models.
 
The right panel illustrates the effect of changing the mass resolution
on \gamn for fixed box sizes of $15h^{-1}$ and $30h^{-1}$ comoving
Mpc.  Increasing the mass resolution decreases the optical depth as
higher resolution simulations resolve smaller haloes.  More gas is
taken from the low density IGM and a smaller value of \gamn is
required to match the observed mean flux.  Simulations with less than
$400^{3}$ gas particles within a $30h^{-1}$ comoving Mpc box do not
achieve adequate numerical convergence.  The difference between \gamn
calculated from the $200^{3}$ and $400^{3}$ simulations is on average
$7$ per cent.  Comparison of the simulation with $200^{3}$ and
$400^{3}$ particles within a $15h^{-1}$ comoving Mpc box shows that a
further increase in resolution reduces \gamn by an extra $5$ per cent.

Unfortunately, $400^{3}$ simulations are too computationally expensive
for the purpose of exploring a large parameter space.  We compromise
by taking our fiducial model to have $200^{3}$ gas particles within a
$15h^{-1}$ comoving Mpc box; the value of \gamn we infer will be
systematically high by around $8$ per cent due to the combined error
from box size and resolution.  Hereafter, all results based on
simulations with $200^{3}$ particles within a $15h^{-1}$ comoving Mpc
box will be reduced by this factor to correct for box size and
resolution.  Note that this correction is likely to be a lower limit
to the true correction required; our highest resolution simulations
may not yet have reached numerical convergence.  We will account for
this uncertainty in our error analysis in section~\ref{sec:errors}.

\section{Scaling relations for $\Gamma_{-12}$ from simulated absorption spectra} \label{sec:scale}

It is implicit in the FGPA scaling relation that the density and
velocity distribution, along with the effective equation of state of
the \Lya absorbers remain unchanged for different values of
$\Omega_{\rm b}$, $h$, $\Omega_{\rm m}$, $T$ and $\Gamma_{-12}$.
Hence, when inferring \gamn from hydrodynamical simulations and
subsequently rescaling with different parameters, it is generally
assumed that:

\begin{equation} \Gamma_{-12} \propto  \Omega_{\rm b}^{2}h^{3}T^{-0.7}\Omega_{\rm m}^{-0.5}  \label{eq:gscale} \end{equation}

\noindent
(e.g. Rauch et al. 1997, McDonald \& Miralda-Escud\'{e} 2001).
However, it is not clear if this relation is an accurate
approximation.  Altering \OM is likely to alter the gas density
distribution, and thermal broadening may produce a deviation from the
$T^{-0.7}$ scaling based on the temperature dependence of the
recombination coefficient.  In addition, the dependence of \gamn on
the polytropic index $\gamma$ of the effective equation of state $T =
T_{0} \Delta^{\gamma -1}$ is unclear.  Although hydrodynamical
simulations predict $\gamma \simeq 1.6$ at low density, if \HeII
reionization occurs at $z \simeq 3$ \citep{b46,b27,b39} the effective
equation of state for the low density gas may become isothermal.
Radiative transfer effects may also raise the temperature of the gas
and blur the monotonic relation between temperature and density
\citep{b50,b47}.  Finally, changing \sig will affect \gamn by changing
the density distribution of the gas.

We test the validity of equation~\ref{eq:gscale} using our suite of
hydrodynamical simulations.  The motivation for this is two-fold:
first, it is important to understand the physical processes which
alter the value of \gamn we infer from simulations; second, we wish to
determine how \gamn scales with the input parameters, allowing us to
place joint constraints upon them.  We do not discuss the scaling with
\OB and $h$, since our analysis revealed that the FGPA scaling
relation holds extremely well for these parameters.

\subsection{The dependence of $\Gamma_{-12}$ on temperature}

 \begin{figure*}
   \centering
\begin{minipage}{180mm}

  \psfig{figure=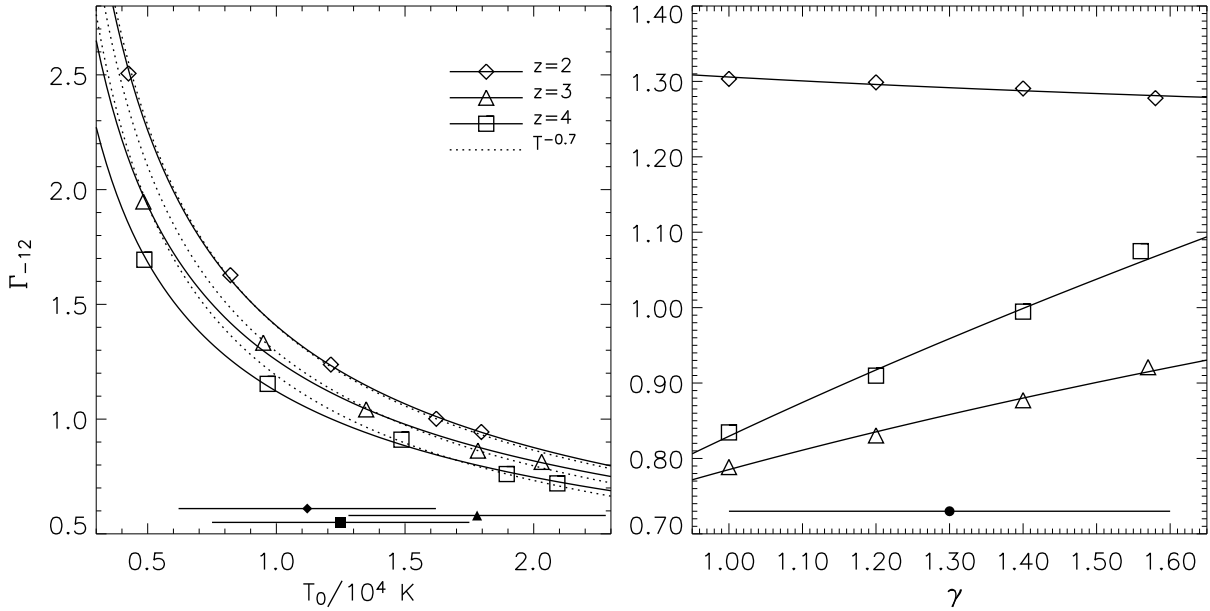,width=1.0\textwidth}
  
\caption{
  {\it Left:} The dependence of $\Gamma_{-12}$ on the gas temperature
  at mean density $T_{0}$ at three different redshifts as indicated on
  the plot. Results are corrected for box size and resolution as
  discussed in section ~\ref{sec:res}.  Solid curves show a least
  squares fit and the dotted curves shows the $T^{-0.7}$ scaling due
  to the temperature dependence of the recombination coefficient.  The
  filled symbols show the fiducial temperatures and their
  uncertainties.  {\it Right:} The dependence of $\Gamma_{-12}$ on the
  index of the temperature density relation $\gamma$.  Results are
  corrected for box size and resolution.  The filled circle shows our
  fiducial value of $\gamma$ and its uncertainty. }
\label{fig:scale}
\end{minipage}
\end{figure*}
\begin{figure*}
  \centering
\begin{minipage}{180mm}

  \psfig{figure=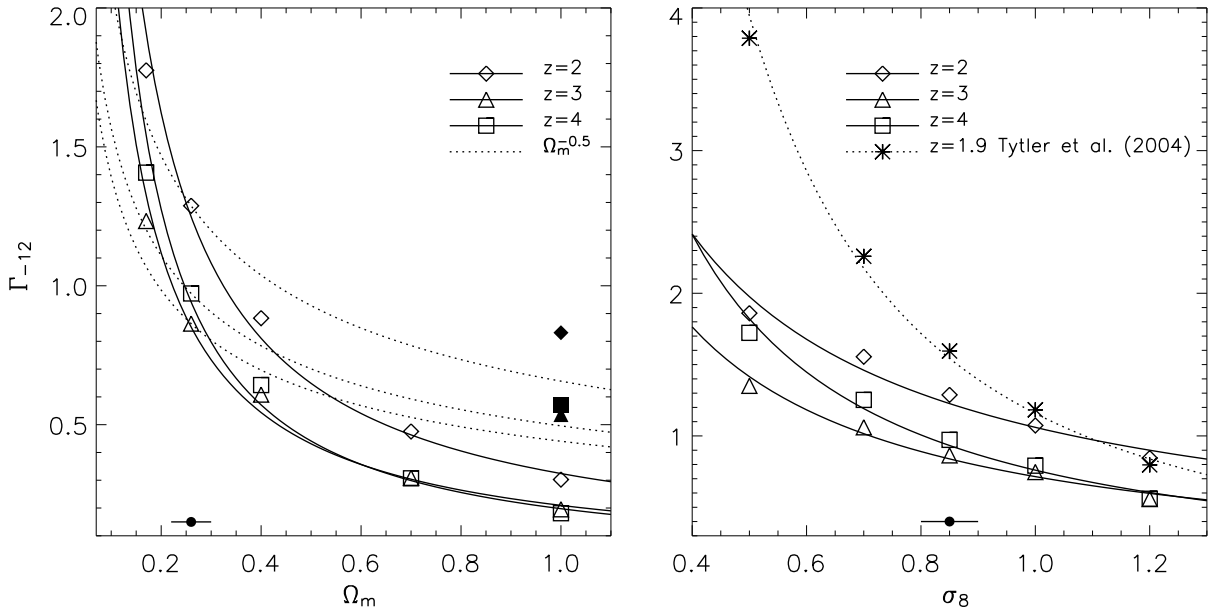,width=1.0\textwidth}
       
\caption{
  {\it Left:} The dependence of the estimated $\Gamma_{-12}$ on
  $\Omega_{\rm m}$ for three different redshifts as indicated on the
  plot.  The solid curves are a least-square fit.  The dotted lines
  show the $\Omega_{\rm m}^{-0.5}$ scaling. Filled points are obtained
  for a model with $\Omega_{\rm m}=1.0$ and the the same r.m.s.
  fluctuation amplitude at a scale of $30$ $\rm kms^{-1}$ as our
  fiducial model with $\Omega_{\rm m}=0.26$. As before the values are
  corrected for box size and resolution. The filled circle shows our
  fiducial value of $\Omega_{\rm m}=0.26$ and its uncertainty.  {\it
    Right:} The dependence of $\Gamma_{-12}$ on $\sigma_{8}$.  The
  dotted curve shows the result of Tytler et al.  at $z=1.9$, assuming
  $\langle F \rangle_{\rm obs} = 0.882 \pm 0.01$.  Results are
  corrected for box size and resolution.  The filled circle shows our
  fiducial value of $\sigma_{8}$ and its uncertainty. }
\label{fig:sigma}
\end{minipage}
\end{figure*}

The left panel of fig.~\ref{fig:scale} shows our test of the FGPA
scaling relation for the gas temperature at mean density using five
simulations with photoheating rates [0.3,1.0,2.0,3.3,4.0] times the
HM96 values.  We expect the ionization rate to decrease as the HM96
photoheating rate is raised; the hydrogen recombination rate scales as
$T^{-0.7}$, so for larger temperatures there is less neutral hydrogen.
We find a modest deviation from the $T^{-0.7}$ scaling (shown as the
dotted curves) at low temperatures, probably due to thermal broadening
of absorption features, an effect which is strongest at $z=4$.  The
fiducial model with its higher gas temperatures will have relatively
broader absorption features, producing more \Lya absorption and thus
requiring a larger value of \gamn than if thermal broadening were
absent.  The FGPA scaling relation for the fiducial model will
slightly over-predict the value of \gamn at lower temperatures.  The
same process also occurs for gas temperatures higher than the fiducial
model; the absorbers are thermally broadened to a greater extent so
the scaling relation under-predicts the ionization rate.  This effect
is much more apparent at higher redshifts when the spatial density of
\Lya absorbers is greater.  Note that the temperature dependence of
the gas pressure may also contribute to the deviation from the
$T^{-0.7}$ scaling.  We perform a least squares fit on the data to
obtain the following scaling for \gamn with temperature:

\begin{equation} \Gamma_{-12} \propto T^{x_{1}(z)} \end{equation}

\noindent
where $x_{1}(z) = [-0.68,-0.62,-0.59]$ at $z=[2,3,4]$.  We shall adopt
these scaling relations (shown as solid curves) from here onwards.

\subsection{The dependence of $\Gamma_{-12}$ on $\gamma$} \label{sec:index}

The dependence of the hydrogen ionization rate on the polytropic index
$\gamma$ is shown in the right panel of fig.~\ref{fig:scale}. The
value of $\gamma$ is varied by artificially rescaling the effective
equation of state of our existing fiducial simulation data by pivoting
the temperature-density relation around the mean gas density
\citep{b8}.  Note that the actual slope of the effective equation of
state predicted by the numerical simulations is $\gamma \simeq 1.6$.
This will not account for any dynamical effects the different
temperatures have on the gas distribution, but it will model thermal
broadening and the change in neutral hydrogen density correctly, thus
providing a reasonable approximation.  Note that the data point with
the largest $\gamma$ is for our simulation of the fiducial model with
no scaling of $\gamma$.  It is clear from fig.~\ref{fig:scale} that
the inferred ionization rate has a non-negligible dependence on
$\gamma$.  At $z=3$ and $z=4$ \gamn increases with $\gamma$.  A
flatter effective equation of state produces hotter gas at densities
less than the cosmic mean with a lower neutral hydrogen fraction.  As
most of the \Lya absorption is produced by this gas, decreasing
$\gamma$ will decrease the amount of absorption and hence lower the
inferred ionization rate.  The dependence of \gamn on $\gamma$ is much
stronger at $z=4$ compared to $z=3$.  However, at $z=2$, the trend is
opposite; the ionization rate decreases with increasing $\gamma$,
albeit with a weak dependence.  This can be explained by considering
the density evolution of the \Lya absorbers.  With decreasing
redshift, the mean gas density decreases.  By $z=2$, the optical depth
in regions of average density and below has become so low that these
contribute very little to the overall effective optical depth. The
absorption is almost entirely due to overdense regions.  Decreasing
the polytropic index will lower the temperature of this gas,
increasing the neutral hydrogen fraction and slightly increasing the
effective optical depth.  The inferred ionization rate increases
slightly with decreasing $\gamma$ as a consequence.  In a similar
manner to before, we perform a least squares fit on the data to obtain
a scaling relation for the ionization rate with the polytropic index
$\gamma$:

\begin{equation} \Gamma_{-12} \propto \gamma^{x_{2}(z)} \end{equation}

\noindent
where $x_{2}(z) = [-0.04,0.34,0.55]$ at $z=[2,3,4]$.  These scaling
relations are shown in fig.~\ref{fig:scale} as the solid curves.

\subsection{The dependence of $\Gamma_{-12}$ on $\Omega_{\rm m}$}\label{sec:om}

The FGPA scaling relation assumes that the ionization rate per atom
scales as $\Omega_{\rm m}^{-0.5}$.  For this to be true, the density
distribution of the neutral hydrogen gas must remain the same for
different values of $\Omega_{\rm m}$.  We test this scaling relation
using five simulations with \OM$=[0.17,0.26,0.4,0.7,1.0]$ and show the
results in the left panel of fig.~\ref{fig:sigma}.  As before, the
plotted values are rescaled to match the fiducial temperatures,
$\gamma=1.3$, corrected for box size and resolution and are shown with
the best fitting curves to the data.  The $\Omega_{\rm m}^{-0.5}$
scaling around the fiducial model is indicated by the dotted lines.
This becomes a poorer fit to the data as \OM is increased, and
overestimates the value of \gamn substantially.  It is worth stressing
that all the models shown in the left panel of fig.~\ref{fig:sigma}
have the same normalisation at a scale of $8h^{-1}$ Mpc for the dark
matter linear power spectrum ($\sigma_8 = 0.85$). In order to
understand why the discrepancy with the expected trend is so big we
performed a new run for the Einstein-de Sitter (M4) model and
normalised the power spectrum to have the same fluctuation amplitude
as the fiducial model at the scale of $30$ $\rm kms^{-1}$, roughly
corresponding to the Jeans length. The filled points in the left panel
of fig.~\ref{fig:sigma} indicate the new values of \gamn we infer; one
can see that these values are in better agreement with the theoretical
curve, showing that \gamn is more sensitive to the r.m.s.
fluctuations on a fixed velocity scale.  This suggests that the r.m.s.
fluctuation amplitude at the Jeans scale is more relevant.

\begin{figure}
  \centering \psfig{figure=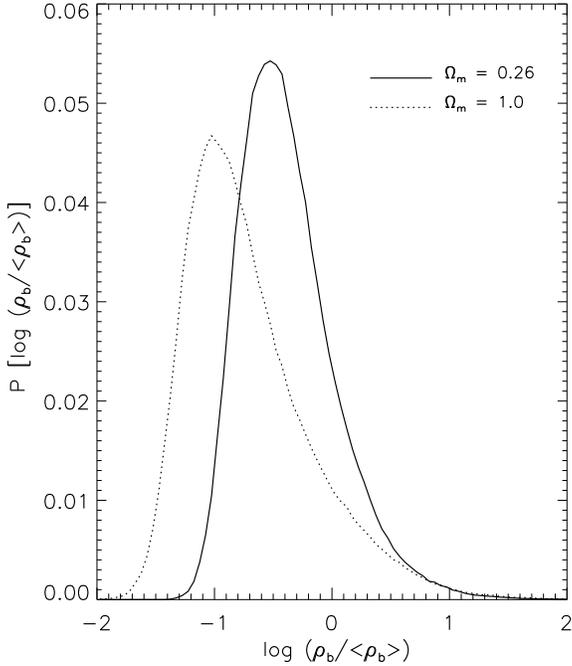,width=0.43\textwidth}

\caption{
  The PDF of the volume weighted gas density as a fraction of the mean
  at $z=2$ for the fiducial \OM$=0.26$ (solid line) and M4 \OM$=1.0$
  (dotted line) models.  The fiducial model has a gas density
  distribution which peaks at a greater amplitude closer to the mean
  density, producing more \Lya absorption at fixed $\Gamma_{-12}$.}
\label{fig:rhoPDF}
\end{figure}

Fig.~\ref{fig:rhoPDF} shows the probability density function (PDF) of
the volume weighted gas density as a fraction of the mean at $z=2$ for
our fiducial model and the Einstein-de Sitter (M4) model.  The density
distribution of the Einstein-de Sitter model peaks at lower density,
explaining the much lower \gamn required to match the observed mean
flux. In the Einstein-de Sitter model a greater proportion of the gas
has fallen into haloes.  Different gas density distributions are not
accounted for in the FGPA scaling relation.  Comparisons of parameters
such as \gamn and \OBh inferred from simulations with different matter
distributions using the FGPA gives misleading results.  A least
squares fit to our data yields

\begin{equation} \Gamma_{-12} \propto \Omega_{m}^{x_{3}(z)}, \end{equation}

\noindent
where $x_{3}(z) = [-1.00,-1.04,-1.16]$ at $z=[2,3,4]$.

\subsection{The dependence of $\Gamma_{-12}$ on $\sigma_{8}$} \label{sec:sig}

The magnitude of $\sigma_{8}$, the mass fluctuation amplitude within
$8h^{-1}$ Mpc spheres, also alters the required \gamn.  For small
$\sigma_{8}$, there is less small scale power and the collapse of
structures is less advanced.  This produces a narrower volume weighted
density PDF with a larger average density.  This decreases the mean
flux of the simulated spectra if the ionization rate is kept fixed.  A
larger \gamn is thus required to reproduce the observed effective
optical depth.

\begin{figure}
  \centering \psfig{figure=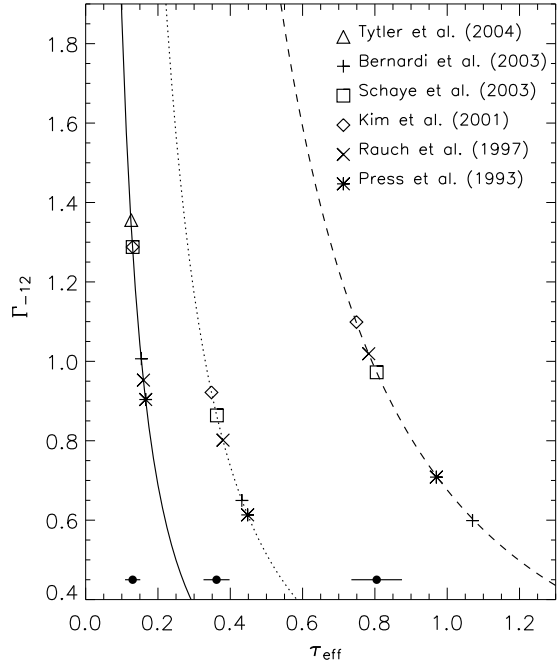,width=0.43\textwidth}
 
\caption{
  The dependence of $\Gamma_{-12}$ on the effective optical depth at
  different redshifts for our fiducial model with our fiducial values
  of $T_{0}$ and $\gamma$.  The symbols indicate the values of
  $\tau_{\rm eff}$ quoted by various authors.  The \teff from Tytler
  et al. (2004) is for $z=1.9$. The filled circles indicate our
  fiducial parameter values, with the estimated uncertainty shown by
  the horizontal line.}
\label{fig:teff}
\end{figure}

The right panel of fig.~\ref{fig:sigma} shows \gamn for
\sig$=[0.5,0.7,0.85,1.0,1.2]$ at redshifts $z=[2,3,4]$.  \citet{b5}
present a similar plot in their fig. 16, where they construct curves
relating \sig and $\gamma_{912} = \Gamma_{-12}/1.329$, normalised to
the predicted ionization rate at $z=1.9$ from \citet{b37}.  The
central curve in their plot gives the \gamn and \sig they require to
reproduce \Fobs$=0.882 \pm 0.01$.  We find a weaker dependence of
\gamn on $\sigma_{8}$. We use similar values for $\Omega_{\rm m}$,
$h$, $\Omega_{\rm b}$, the UV heating rate, and the slightly lower
value of $\tau_{\rm eff}$, which Tytler et al. use, will change our
results only by a small amount (see section~\ref{sec:teff}).  The
small difference in redshift should also not alter the results
significantly, and the resolution and box size of our simulations will
also not account for the difference; increasing these systematically
lowers $\Gamma_{-12}$.  The most likely reason for this discrepancy
are differences in the modelling techniques for our respective
cosmological simulations \citep{b68}.  Tytler et al.  use the Eulerian
code ENZO \citep{b44}. This models the gas component using a piecewise
parabolic method, whereas we use a Lagrangian SPH code.  A least
squares fit to our results gives:

\begin{equation} \Gamma_{-12} \propto \sigma_{8}^{x_{4}(z)} \end{equation}

\noindent
where $x_{4}(z) = [-0.90,-0.99,-1.26]$ at $z=[2,3,4]$.

\subsection{The dependence of $\Gamma_{-12}$ on $\tau_{\rm eff}$} \label{sec:teff}

To determine the intensity and evolution of \gamn using the FGPA we
must assume a value for the effective optical depth, $\tau_{\rm eff} =
- \rm ln \langle F \rangle_{\rm obs}$. Measurements of the effective
optical depths have been made by a number of authors (e.g Press,
Rybicki \& Schneider 1992, Kim et al. 2001, Songalia \& Cowie 2002,
Bernardi et al. 2003, Schaye et al.  2003).  The values obtained from
low-resolution, low-S/N data by Press, Rybicki \& Schneider and
Bernardi et al. appear to be systematically higher than those obtained
by the other groups, mainly from high-resolution data (see Viel,
Haehnelt \& Springel 2004 for a discussion).  The differences are
primarily due to the systematic uncertainties in the continuum fits.
Tytler et al.  make an analysis of continuum fitting errors by
comparing the true and fitted continua for sets of artificial spectra.
They find that they can obtain a fit which is on average good to
within 1-2 per cent, except for very low S/N spectra.  Some agreement
is emerging that the lower values from the higher quality data are
more appropriate.  Statistical errors from the presence of metal
absorption features and high column density absorbers also contribute
to the uncertainty in \teff \citep{b69}.

Fig.~\ref{fig:teff} shows the value of \gamn we infer from our
fiducial model at $z=[2,3,4]$ by varying the effective optical depth,
and hence mean observed flux, to which the simulated optical depths
are scaled to match.  At $z=2$, a change in \teff of $0.05$, which is
consistent with the variation in estimates, results in almost a factor
of two difference in the required hydrogen ionization rate.  An
accurate estimate of \teff and the errors associated with the
measurement are required to determine $\Gamma_{-12}$; at $z=2$ even a
small change in the assumed \teff will change the inferred ionization
rate appreciably.  The best fitting curve to the data is:

\begin{equation} \Gamma_{-12} \propto \tau_{\rm eff}^{x_{5}(z)} \end{equation}

\noindent
where $x_{5}(z)=[-1.44,-1.61,-1.68]$ at $z=[2,3,4]$.

\section{Comparison with previous estimates from simulations} \label{sec:comp}
\subsection{Hydrodynamical codes}

In this section we shall make a direct comparison between our data and
the results of other authors.  Firstly, we consider two simulations
which are similar to the \LCDM and SCDM models analysed by Rauch et
al. (1997).  The simulations have identical cosmological parameters,
box size and resolution.  Our \LCDM model differs from that analysed
by Rauch et al., since we use an SPH code rather than an Eulerian grid
code.  We try to approximate the different temperatures of the models
by increasing the photoheating rate to $3.3$ times the HM96 value for
our \LCDM simulation.  This results in similar temperatures as those
given by Rauch et al. at $z=2$ (this is only redshift for which the
temperature was published).  The optical depths of our artificial
spectra are scaled to match the mean flux used by Rauch et al.,
corrected for continuum fitting: \Fobs$=[0.848,0.670,0.414]$ and
$[0.846,0.655,0.383]$ at $z=[2,3,4]$ for the \LCDM and SCDM models
respectively.  Mimicking Rauch et al., we rescale \gamn by a factor
$(\Omega_{\rm b}h^{2})^{2}$, setting \OBh$=0.024$.  Unlike in the rest
of the paper, we have not corrected our data for numerical effects or
rescaled the temperatures to match our fiducial values of $T_{0}$ and
$\gamma$.

\begin{figure}
  \centering

  \psfig{figure=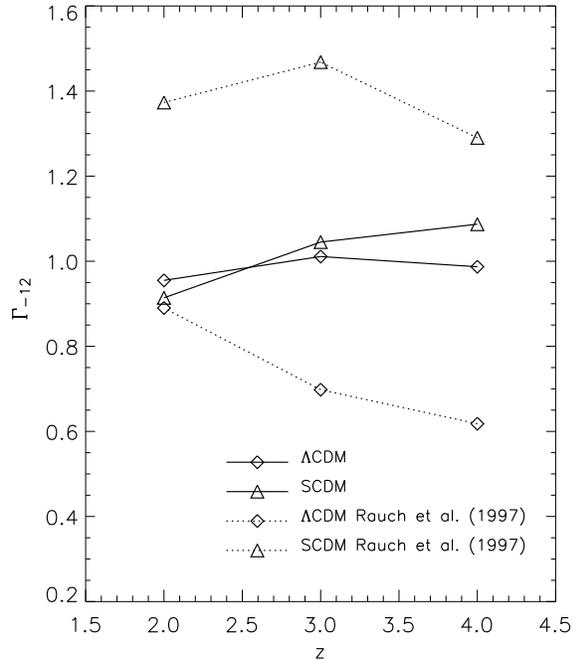,width=0.43\textwidth}
       
\caption{Comparison of the ionization rate inferred from GADGET-2  to that inferred from the simulations
  used in Rauch et al. (1997).  Diamonds and triangles represent the
  $\Lambda$CDM and SCDM models, respectively.  The GADGET-2 results
  are shown with solid lines while the results from Rauch et al. are
  shown with dotted lines.}
\label{fig:rauch}
\end{figure}

Fig.~\ref{fig:rauch} shows the redshift dependence of \gamn inferred
from the \LCDM and SCDM models as diamonds and triangles respectively.
The solid lines give our data and the Rauch et al. data are plotted as
dotted lines.  Rauch et al. conclude the large difference between
\gamn obtained using the \LCDM and SCDM models is primarily due to
different gas temperatures in the respective simulations.  The gas
temperature at mean density in their SCDM model is about a half of
that in the \LCDM model at $z=2$.  We suggest the values of
$\Omega_{\rm m}$, \sig and $\gamma$ used in each of the simulations
will also alter $\Gamma_{-12}$.  The different amounts by which the
optical depths of the artificial spectra are scaled will also produce
a small difference in $\Gamma_{-12}$, but this change will be
negligible compared to the effect of the other parameters.  In our
simulations the smaller temperature of the SCDM simulation counteracts
the effect of different \OM and \sig and we find rather similar values
for both simulations. The values for our \LCDM simulation are larger
than those in Rauch et al.  while those for the SCDM simulation are
smaller.  Unfortunately the only information about the thermal state
of the gas provided by Rauch et al. is at $z=2$ which makes a accurate
comparison difficult.  Although we attempt to match the gas
temperature in our models at $z=2$ to those given by Rauch et al.,
this agreement is probably not good at higher redshifts, particularly
for the \LCDM model.  The gas temperatures of the Rauch et al. models
result from different reionization histories instead of artificially
increased photoheating rates.  The UV background of their \LCDM model
ionizes \HeII at $z\simeq 3$, which may lower the polytropic index
$\gamma$. The $\gamma$ in the Rauch et al. \LCDM simulation may thus
have been lower than that in our simulations.  This would make the
discrepancy smaller.

\citet{b41} utilise a hydrodynamical simulation based on a modified
version of the SPH code HYDRA \citep{b51} to examine the metallicity
of the IGM.  Their model consists of a periodic box of $12h^{-1}$
comoving Mpc with $256^{3}$ dark matter particles and $256^{3}$ gas
particles.  The parameters they adopt are \OM$=0.3$, \OBh$=0.019$,
$h=0.65$, \sig$=0.9$ and $n=1.0$ with a UV background designed to
reproduce the effective equation of state constrained by \citet{b27}.
The optical depths of the artificial \Lya spectra are scaled to match
the same mean flux as ours.  They find the \gamn required to match the
mean flux of the \Lya forest at $z=[2,3,4]$ are $\Gamma_{-12} =
[0.87,0.54,0.36]$.  Rescaling our fiducial model values of
$\Gamma_{-12} = [1.29,0.86,0.97]$ to match the cosmological parameters
used by Schaye et al. reduces our values by around $40$ per cent to
$\Gamma_{-12} =[0.80,0.53,0.57]$.  We remove the correction for box
size and resolution from these values to allow a fair comparison.
These are consistent with \citet{b41}, allowing for the unknown
differences between the exact thermal state of the low density gas in
the two simulations.  Theuns, Schaye \& Haehnelt (2000) also run
simulations using HYDRA to study the broadening of \Lya absorption
lines in QSO spectra.  Their reference model consists of $64^{3}$ dark
matter particles and $64^{3}$ gas particles within a periodic box of
$2.5h^{-1}$ comoving Mpc, with the same cosmological parameters used
by Schaye et al. and the UV background specified by Haardt \& Madau
(1996).  This produces a temperature of $T_{0} \simeq 11750$ $\rm K$
at mean cosmic density.  To reproduce an effective optical depth of
$\tau_{\rm eff}=0.33$ at $z=3$ they require $\Gamma_{-12}=0.87$, based
on the value of $A$ quoted in their paper.  Scaling our fiducial data
to match their parameters reduces our value by $10$ per cent to
$\Gamma_{-12}=0.79$.  Although the mass resolution of our simulations
is comparable, we obtain a smaller value because of our larger
simulation box size.

Most recently, \citet{b5} made a detailed analysis of the \Lya forest
at $z=1.9$, using a set of $77$ moderate resolution QSO spectra and
compared it to a large hydrodynamical simulation run with the Eulerian
code ENZO.  Including uncertainties on the value of $\sigma_{8}$, \OB
and $\tau_{\rm eff}$, they find a joint constraint of \gamn$=1.44
\pm0.36$.  They use a hydrodynamical simulation on a $1024^{3}$ grid
in a $54.528h^{-1}$ comoving Mpc box, with \OM=$0.27$, \OBh$=0.022$,
$h=0.71$, \sig$=0.9$ and $n=1.0$ and a UV background specified by
\citet{b37} with an artificially increased \HeII photo-heating rate.
Our fiducial result at $z=2$ rescaled to match the cosmological
parameters and $\tau_{\rm eff}=0.126$ at $z=1.9$ assumed by Tytler et
al. gives \gamn$=1.17$.  This value is in reasonable agreement with
the Tytler et al. result, again allowing for differences between the
exact gas temperatures and numerical method.  Note, however that
Tytler et al. found a different scaling of \gamn with $\sigma_{8}$
(see section~\ref{sec:sig} for details).

\subsection{Pseudo-hydrodynamical codes}

Several authors have also considered this problem using
pseudo-hydrodynamical codes (e.g Gnedin \& Hui, Gnedin 1998, Meiksin
\& White 2003, 2004).  These approximate methods sacrifice accuracy
for speed, allowing a much larger parameter space to be explored.
Providing the accuracy of these approximations and the parameter
ranges they model effectively are well controlled, they can provide an
alternative method for modelling the IGM. \citet{b53} concludes that
observational constraints on the column density distribution of the
\Lya forest are consistent with the intensity of the UV background
remaining constant in the redshift range $z \sim 2-4$.  \citet{b55},
in a study of UV background fluctuations, find the mean flux of the
IGM is reproduced with
$\Gamma_{-12}=[0.88^{+0.14}_{-0.12},0.76^{+0.12}_{-0.11}]$ at
$z=[3,4]$.  They adopt \OM=$0.3$, \OBh$=0.020$, $h=0.70$, \sig$=1.0$
and $n=1.0$, and impose and effective equation of state with
$T_{0}=20000$ $\rm K$ and $\gamma=1.5$.  \citet{b60} find that the
spectral properties of the \Lya forest obtained from their code are
reproduced to within $10$ per cent of those found using fully
hydrodynamical simulations.  These results can thus be considered to
be consistent with ours.

\section{The Metagalactic Hydrogen Ionization Rate}\label{sec:results}
 
\subsection{The error budget}\label{sec:errors}

\begin{table} 
\centering
\caption{Fiducial parameter values and estimates of their uncertainty.  The values listed for gas temperature at mean density $T_{0}$ and the effective optical depth \teff are for $z=[2,3,4]$.  }

\begin{tabular}{c|c}
  \hline
    Parameter & Fiducial values and uncertainties\\  
    \hline
    $T_{0}$          & [11200, 17800, 12500]$\pm$5000 K   \\
    \OM          & 0.26$\pm$0.04 \\
    \teff        & [0.130$\pm$0.021, 0.362$\pm$0.036, 0.805$\pm$0.070]  \\
    $\gamma$     & 1.3$\pm$0.3 \\
    \OBh         & 0.024$\pm$0.001\\
    \sig         & 0.85$\pm$0.05  \\
     $h$          & 0.72$\pm$0.04   \\
    \hline  
    \label{tab:fiducial}
  \end{tabular}

\end{table}

\begin{table} 
\centering
\caption{
  Percentage error budget for $\Gamma_{-12}$ from estimates of various cosmological and astrophysical parameters, listed approximately in order of importance.  The total error is obtained by adding the individual errors in quadrature.}

\begin{tabular}{c|c|c|c|c}
  \hline
    Parameter & $z=2.00$ & $z=3.00$ & $z=4.00$ \\  
    \hline
    $T$           & $^{+50}_{-22}\%$ & $^{+23}_{-14}\%$ & $^{+35}_{-18}\%$   \\
    & & & \\
    \OM           & $^{+18}_{-13}\%$ & $^{+19}_{-14}\%$ & $^{+21}_{-15}\%$   \\
    & & & \\
    \teff         & $^{+29}_{-19}\%$ & $^{+18}_{-14}\%$ & $^{+17}_{-13}\%$   \\
    & & & \\
    Numerical     &  $\pm$10$\%$ &  $\pm$10$\%$ &  $\pm$10$\%$  \\  
    & & & \\
    $\gamma$      & $\pm$1$\%$ & $^{+7}_{-9}\%$ & $^{+12}_{-13}\%$   \\
    & & & \\
    \OBh          & $^{+9}_{-8}\%$ & $^{+9}_{-8}\%$ & $^{+9}_{-8}\%$   \\
    & & & \\
    \sig          & $^{+6}_{-5}\%$ & $\pm$6$\%$ & $^{+8}_{-7}\%$   \\
    & & & \\
    $h$           &  $\pm$6$\%$ &  $\pm$6$\%$ &  $\pm$6$\%$ \\ 
    \hline
    Total & $^{+62}_{-36}\%$ & $^{+39}_{-30}\%$ & $^{+49}_{-34}\%$   \\
    \hline  
    \label{tab:errors}
  \end{tabular}

\end{table}
 
\begin{figure*}
  \centering
\begin{minipage}{180mm}

  \psfig{figure=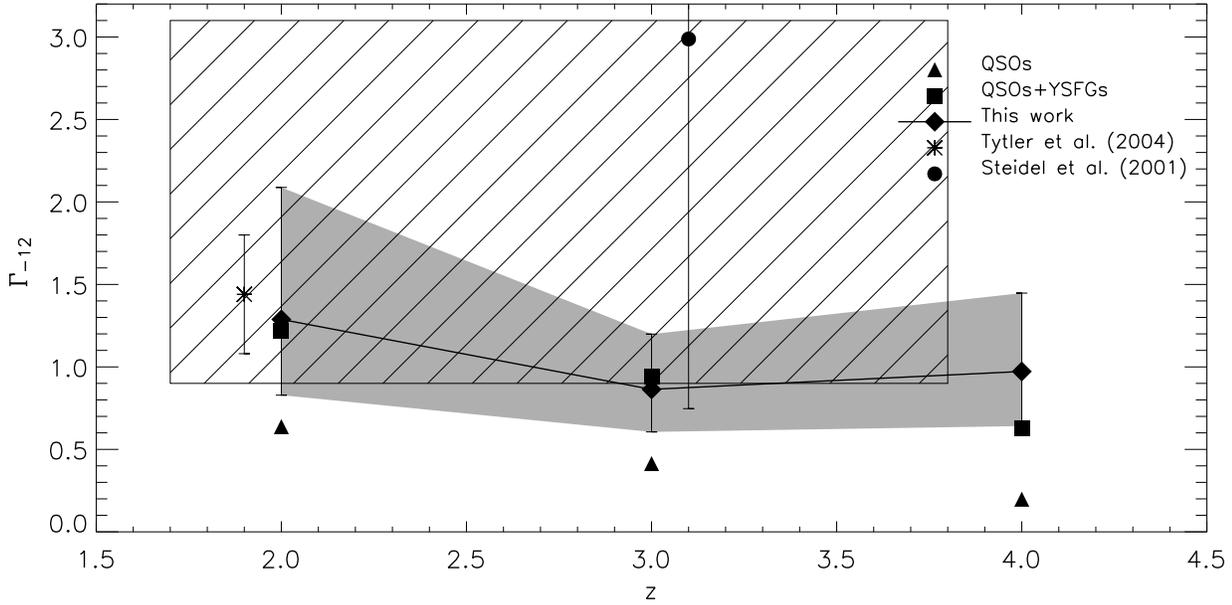,width=1.0\textwidth}
       
\caption{
  Comparison of our best estimate for $\Gamma_{-12}$ at $z=[2,3,4]$
  with constraints from observations.  Our data are plotted with
  filled diamonds, and the grey shaded area shows the error bounds.
  The filled squares and triangles show the estimated contribution to
  the metagalactic ionization rate from QSOs+galaxies and QSOs alone,
  based on estimates from the updated model of \protect\citet{b37}
  including UV photons from re-processing by the IGM.  The hatched box
  gives the constraint on \gamn from the proximity effect
  \protect\citep{b24} and the estimate from Lyman-break galaxies
  assuming a global spectral index of $\alpha=1.8$ is plotted with a
  filled circle \protect\citep{b23}.  The data point has been offset
  from $z=3$ for clarity, and the upper error limit is not shown.  The
  star shows the best estimate of \gamn at $z=1.9$ from
  \protect\citet{b5}, including their errors from the uncertainty in
  $\tau_{\rm eff}$, \sig and $\Omega_{\rm b}$.}
\label{fig:HM}
\end{minipage}
\end{figure*}

We now proceed to place joint constraints on the metagalactic hydrogen
ionization rate using independent estimates for several cosmological
and astrophysical parameters.  We adopt the values listed in
Table~\ref{tab:fiducial}.  The values of $T_{0}$ and $\gamma$ are
based on the results of Schaye et al. (2000) and we assume the \Lya
effective optical depth advocated by Schaye et al. (2003).  The
cosmological parameters are consistent with the results of Spergel et
al. (2003), based on their analysis of the cosmic microwave background
anisotropy.  Using our scaling relations, we estimate the error on the
values of \gamn from our fiducial simulation, corrected for box size
and resolution.  The total error is calculated by adding the
individual contributions in quadrature, with the percentage error on
\gamn contributed by each parameter listed in Table~\ref{tab:errors}.
We make the conservative assumption of an additional numerical
uncertainty of $10$ per cent to take into account numerical
convergence and the modest differences with numerical simulations of
other authors.  The total percentage errors on \gamn are
$[\pm^{62}_{36},\pm^{39}_{30},\pm^{49}_{34}]$ per cent at $z=[2,3,4]$,
with the largest contribution coming from the uncertainty in the gas
temperature.  There is also a large contribution from the uncertainty
in the effective optical depth, especially at $z=2$.  Interestingly,
the uncertainty in \OM also gives a substantial contribution to the
total error budget.  This is primarily due to the sensitivity of \gamn
on \OM due to changes in the r.m.s.  fluctuations of the gas density
at the Jeans scale.  Other uncertainties are less important, in
particular $\Omega_{\rm b}h^{2}$, \sig and $h$.  This contrasts with
the conclusion of Tytler et al., who find \sig is the biggest
contribution to the uncertainty on $\Gamma_{-12}$.  However they find
a stronger dependence of \gamn on \sig than we do, and they do not
investigate the contribution of uncertainties in \OM or $T$ to this
error. Note also that inhomogeneities in the UV background may
increase estimates of \gamn by up to twenty per cent if the UV
background is dominated by emission of QSOs (Gnedin \& Hamilton 2002;
Croft 2004; Meiksin \& White 2004). We will conclude in section 5.2.
that QSOs contribute less than half of the ionizing UV background. We
will therefore neglect a possible error due to inhomogeneities in the
UV background.  Using our fiducial model, we find the \Lya effective
optical depth of the IGM at $z=[2,3,4]$ is reproduced by
\gamn$=[1.29\pm^{0.80}_{0.46},0.86\pm^{0.34}_{0.26},0.97\pm^{0.48}_{0.33}]$.

\subsection{Comparison with other observational estimates}
 
We compare our results with other observational estimates of \gamn in
fig.~\ref{fig:HM}.  Our estimates are plotted as filled diamonds, with
the error bound shown by the dark grey shaded region.  The filled
triangles show the metagalactic hydrogen ionization rate computed from
the updated UV background model of \citet{b37}, hereafter MHR99
(Francesco Haardt, private communication).  The values quoted here are
smaller by more than a factor of two than those presented in the work
of HM96, primarily due to different assumptions about the emissivity
of the ionizing sources, and to a lesser extent upon the assumed
cosmology.  The rates are based on the QSO optical luminosity function
of Boyle et al. (2000), with a redshift evolution similar to that of
MHR99, aside from a steeper evolution for $z>3$ to fit data from the
SDSS survey (Fan et al. 2001).  These values are computed assuming the
current cosmological concordance model, and also include the
contribution of diffuse radiation from re-processing by the IGM ({\it
  e.g.} HM96).  Our data appear to be inconsistent with the IGM being
kept ionized by QSOs at $z\leq4$.  For comparison, the filled squares
give \gamn calculated using the QSO rate above, plus an additional
source in the form of Young Star Forming Galaxies (YSFGs).  The YSFG
rates are based on the models of Bruzual \& Charlot (1993), assuming a
Salpeter IMF, a constant star formation rate, an age of $500$ Myr, $Z
= 0.2Z_{\odot}$ and a star formation history as in MHR99.  The very
uncertain escape fraction is assumed to be $f_{\rm esc}=0.1$.  Note
that at low redshift the observed value is less than half this
\citep{b66,b65}.

These ionization rates are in good agreement with our results,
suggesting that a substantial contribution from galaxies appears to be
required at all redshifts. Note, however that there are still
substantial uncertainties in the estimates of the ionization rate due
to QSOs and galaxies.  Hunt et al. (2004), Croom et al. (2004) and
Barger et al. (2004) all suggest that the faint end slope of the
luminosity function of optically and UV bright QSOs is shallower than
the extrapolation from the Boyle et al. data, which may further lower
the expected contribution by QSOs.  On the other hand Scott et al.
(2004) find that the UV spectrum of low luminosity QSOs is
substantially harder than that of bright QSOs which may somewhat
increase their contribution to the UV background.  The main
uncertainty in the estimate for the ionization rate due to the UV
emission of galaxies is the escape fraction of ionizing photons.
However, the improved estimates for the QSO emissivity are so low that
there should now be little doubt that the UV background cannot be
dominated by QSOs over the full redshift range $z<2<4$.  Aguirre et
al. (2004) come to similar conclusions in their study of the
\SiIV/\CIV ratio of the associated metal absorption in \Lya forest
spectra.

Constraints on \gamn from the proximity effect measurements of
\citet{b24} are shown by the hatched box, based on a set of 99
moderate resolution QSO spectra.  Similar results have been obtained
by Cooke et al. (1997).  Allowance has been made for variation in the
spectral indices of QSOs and differences in systemic redshifts
measured from the \OIII or \MgII line and \Lya emission.  Our results
are consistent with the lower end of these estimates.  The filled
circle gives the \gamn we calculate assuming a spectral index of
$\alpha=1.8$ using the metagalactic ionizing radiation intensity,
$J_{\nu}(912$ $\rm \r{A})$, inferred from Lyman-break galaxies (LBGs)
\citep{b23}.  Steidel et al. quote $J_{\nu}(912$ $\rm \r{A}) \simeq
1.2 \pm 0.3 \times 10^{-21}$ $\rm erg s^{-1} cm^{-2} Hz^{-1} sr^{-1}$
at $z\sim 3$ based on observations of the bluest 25 per cent of LBGs.
We offset the data point from $z=3$ for clarity, and the upper error
limit extends beyond the range of the plot.  We estimate the maximum
error on this result to be a factor of $4$ to allow for the
possibility that the $75$ per cent of LBGs which are redder may
contribute negligibly to the UV background.  Note that Giallongo et
al. (2002) found an upper limit to the escape fraction from two bright
LBGs which is about a factor four lower than that measured by Steidel
et al..  There are further uncertainties due to the mean free path of
ionizing photons assumed by Steidel et al. and the spectral index that
we assumed.

Finally we also plot the value of \gamn inferred by \citet{b5} in
their study of the \Lya forest at $z=1.9$.  The error bars include the
uncertainties in $\tau_{\rm eff}$, \sig and $\Omega_{\rm b}$, based on
scaling from their hydrodynamical simulations.

\section{Conclusions}\label{sec:cons}

We have examined the dependence of the ionization rate per hydrogen
atom on cosmological and astrophysical parameters using hydrodynamical
simulations.  This is the first time all these parameters have been
studied using fully hydrodynamical SPH simulations with high
resolution in large boxes.  We find that the gas temperature, the
effective optical depth \teff and the mass fraction \OM have a marked
effect upon the hydrogen ionization rate, in addition to the baryon
fraction $\Omega_{\rm b}$, $\sigma_{8}$, $\gamma$ and the Hubble
constant.  We find that \gamn scales around our fiducial model as

\begin{equation} \Gamma_{-12} \propto \Omega_{\rm b}^{2}h^{3}T^{x_{1}(z)}\gamma^{x_{2}(z)}\Omega_{\rm m}^{x_{3}(z)}\sigma_{8}^{x_{4}(z)}\tau_{\rm eff}^{x_{5}(z)},  \label{eq:totscale} \end{equation}

\noindent
where we tabulate the indices of equation $(10)$ in
Table~\ref{tab:param3}.  We stress that this scaling relation is
likely to be somewhat model dependent and should not be applied to
models with parameters very different from our fiducial model without
further checks.  However, it does provide a clear picture of the
degeneracies which exist between \gamn and other parameters within our
simulations, and is independent of assumptions about the gas
distribution, effective equation of state and ionized gas fraction of
the low density IGM.  In particular, the \Lya optical depth at fixed
r.m.s. fluctuation amplitude \sig is more strongly dependent on \OM
than the FGPA suggests due to changes in the gas density distribution.
Thermal broadening also produces a deviation from the FGPA scaling for
temperature, although this change is much less dramatic.
Consequently, we urge caution when using the FGPA scaling relation,
especially if comparing models with differing $\Omega_{\rm m}$.  We
find good agreement with estimates of \gamn from the \Lya forest
opacity by other authors if differences in the assumed cosmological,
astrophysical and numerical parameters are taken into account.

\begin{table} 
\centering
  \caption{The redshift dependent indices from our scaling relation of $\Gamma_{-12}$ with several cosmological and astrophysical parameters.}
  \begin{tabular}{c|c|c|c|c|c}

    \hline
         & $T$ & $\gamma$ & $\Omega_{\rm m}$ & $\sigma_{\rm 8}$ & $\tau_{\rm eff}$ \\
    \hline
    \hline
    $z$  & $x_{1}(z)$ & $x_{2}(z)$ & $x_{3}(z)$ & $x_{4}(z)$ & $x_{5}(z)$ \\        
  \hline
     2   & -0.68 & -0.04 & -1.00 & -0.90 & -1.44 \\
     3   & -0.62 &  0.34 & -1.04 & -0.99 & -1.61 \\
     4   & -0.59 &  0.55 & -1.16 & -1.26 & -1.68 \\
  \hline
  
  \label{tab:param3}
\end{tabular}
\end{table}

In recent years, compelling observational evidence has led to the
acceptance of a standard cosmological model which is flat, has low
matter density and a substantial contribution of dark energy to the
total energy density. Within this 'concordance' cosmological model the
current generation of hydrodynamical simulations predict values for
the metagalactic hydrogen ionization rate, required to reproduce the
effective \Lya optical depth of the IGM in the range $z=2-4$, which
are about a factor of four larger than those in an Einstein-de Sitter
model with the same r.m.s. density fluctuation amplitude $\sigma_{8}$.
The ionization rates estimated from the \Lya forest opacity are more
than a factor two larger than estimates from the integrated flux of
optically/UV bright observed QSOs alone.  This discrepancy increases
with increasing redshift.  We confirm the findings of Tytler et al.
(2004) at $z \sim 1.9$ that the estimated ionization rates from
simulations of a \LCDM concordance model are in reasonable agreement
with the estimates of the integrated ionizing flux from observed QSOs
plus a significant contribution from galaxies as in the model of
MHR99. A model where the UV background is dominated by emission from
QSOs appears to be inconsistent with the \Lya forest data in the
complete redshift range $2<z<4$. The estimates of the ionization rate
are also in agreement with the lower end of the range of estimates
from the proximity effect.

There are still many issues which need to be resolved.  In particular,
better constraints on the thermal history of the IGM are required.
Additional physics such as radiative transfer, galactic feedback and
metal enrichment may need to be incorporated into simulations in a
more realistic fashion.  However, our estimate of the errors on \gamn
most likely accounts for the modest changes expected from these
processes, suggesting we have obtained a consistent constraint on the
metagalactic hydrogen ionization rate.

\section*{Acknowledgements}

We are grateful to Francesco Haardt for making his updated UV
background model available to us.  This research was conducted in
cooperation with SGI/Intel utilising the Altix 3700 supercomputer
COSMOS at the Department of Applied Mathematics and Theoretical
Physics in Cambridge and on the Sun Linux cluster at the Institute of
Astronomy in Cambridge.  COSMOS is a UK-CCC facility which is
supported by HEFCE and PPARC.  We also acknowledge support from the
European Community Research and Training Network ``The Physics of the
Intergalactic Medium''.  This research was supported in part by the
National Science Foundation under Grant No. PHY99-07949.  JSB, MGH and
MV thank PPARC for financial support.  We also thank the referee, Joop
Schaye, for a helpful report.

\label{lastpage}


\begin{thebibliography}{99} 
  
\bibitem[\protect\citeauthoryear{Abel et al.}{1997}]{b31} Abel T.,
  Anninos P., Zhang Y., Norman M.L., 1997, NewA, 2, 181
  
\bibitem[\protect\citeauthoryear{Abel \& Haehnelt}{1999}]{b50} Abel
  T., Haehnelt M.G., 1999, ApJ, 520, L13
  
\bibitem[\protect\citeauthoryear{Aguirre et al.}{2004}]{b87} Aguirre
  A., Schaye J., Kim T.-S., Theuns T., Rauch M., Sargent W.L.W., 2004,
  ApJ, 602, 38
  
\bibitem[\protect\citeauthoryear{Barger et al.}{2004}]{b88} Barger
  A.J., Cowie L.L., Mushotzky R.F., Yang Y., Wang W.-H., Steffen A.T.,
  Capak P., 2004, preprint (astro-ph/0410527)
 
\bibitem[\protect\citeauthoryear{Bajtlik, Duncan \&
    Ostriker}{1988}]{b63} Bajtlik S., Duncan R.C., Ostriker J.P.,
  1988, ApJ, 327, 570
  
\bibitem[\protect\citeauthoryear{Bechtold}{1994}]{b62} Bechtold J.,
  1994, ApJS, 91, 1
  
\bibitem[\protect\citeauthoryear{Bernardi et al.}{2003}]{b39} Bernardi
  M. et al., 2003, AJ, 125, 32
  
\bibitem[\protect\citeauthoryear{Bi, B\"{o}rner \& Chu}{1992}]{b95} Bi
  H.G., B\"{o}rner G., Chu Y., 1992, A\&A, 266, 1
  
\bibitem[\protect\citeauthoryear{Bi \& Davidson}{1997}]{b96} Bi H.G.,
  Davidson A.F., 1997, ApJ, 479, 523
  
\bibitem[\protect\citeauthoryear{Bolton, Meiksin \& White}{2004}]{b47}
  Bolton J., Meiksin A., White M., 2004, MNRAS, 348, L43
  
\bibitem[\protect\citeauthoryear{Boyle et al.}{2000}]{b72} Boyle B.J.,
  Shanks T., Croom S.M., Smith R.J., Miller L., Loaring N., Heymans
  C., 2000, MNRAS, 317, 1014
  
\bibitem[\protect\citeauthoryear{Bruscoli et al.}{2003}]{b70} Bruscoli
  M., et al., 2003, MNRAS, 343, 51
  
\bibitem[\protect\citeauthoryear{Bruzual \& Charlot}{1993}]{b74}
  Bruzual A.G., Charlot S., 1993, ApJ, 405, 538
  
\bibitem[\protect\citeauthoryear{Bryan \& Machacek}{2000}]{b28} Bryan
  G.L., Machacek M.E., 2000, ApJ, 534, 57
  
\bibitem[\protect\citeauthoryear{Cen}{1992}]{b13} Cen R., 1992, ApJS,
  78, 341
  
\bibitem[\protect\citeauthoryear{Cooke, Espey \& Carswell}{1997}]{b64}
  Cooke A.J., Espey B., Carswell B., 1997, MNRAS, 284, 552
 
\bibitem[\protect\citeauthoryear{Couchman, Thomas \&
    Pearce}{1995}]{b51} Couchman H.M.P., Thomas P.A., Pearce F.R.,
  1995, ApJ, 452, 797
  
\bibitem[\protect\citeauthoryear{Croft}{2004}]{b81} Croft R.A.C.,
  2004, ApJ, 610, 642
  
\bibitem[\protect\citeauthoryear{Croom et al.}{2004}]{b90} Croom S.M.,
  Smith R.J., Boyle B.J., Shanks T., Miller L., Outram P.J., Loaring
  N.S., 2004, MNRAS, 349, 1397
  
\bibitem[\protect\citeauthoryear{Desjacques et al.}{2004}]{b35}
  Desjacques V., Nusser A., Haehnelt M.G., Stoehr F., 2004, MNRAS,
  350, 879
  
\bibitem[\protect\citeauthoryear{Eisenstein \& Hu}{1999}]{b30}
  Eisenstein D.J., Hu W., 1999, ApJ, 511, 5
  
\bibitem[\protect\citeauthoryear{Fan et al.}{2001}]{b73} Fan X., et
  al., 2001, ApJ, 122, 2833
  
\bibitem[\protect\citeauthoryear{Ferland et al.}{1992}]{b92} Ferland
  G.J., Peterson B.M., Horne K., Welsh W.F., Nahar S.N., 1992, ApJ,
  387, 95
  
\bibitem[\protect\citeauthoryear{Fern\'{a}ndez-Soto, Lanzetta \&
    Chen}{2003}]{b65} Fern\'{a}ndez-Soto A., Lanzetta K.M., Chen
  H.-W., 2003, MNRAS, 342, 1215

  
\bibitem[\protect\citeauthoryear{Giallongo et al.}{2002}]{b85}
  Giallongo E., Cristiani S., D'Odorico S., Fontana A., 2002, ApJ,
  568, L9

  
\bibitem[\protect\citeauthoryear{Gnedin}{1998}]{b53} Gnedin N.Y.,
  1998, MNRAS, 299, 392
  
\bibitem[\protect\citeauthoryear{Gnedin \& Hamilton}{2002}]{b82}
  Gnedin N.Y., Hamilton A.J.S., 2002, MNRAS, 334, 107

  
\bibitem[\protect\citeauthoryear{Gnedin \& Hui}{1998}]{b67} Gnedin
  N.Y., Hui L., 1998, MNRAS, 296, 44
  
\bibitem[\protect\citeauthoryear{Gunn \& Peterson}{1965}]{b17} Gunn
  J.E., Peterson B.A., 1965, ApJ, 142, 1633
  
\bibitem[\protect\citeauthoryear{Haardt \& Madau}{1996}]{b22} Haardt
  F., Madau P., 1996, ApJ, 461, 20 (HM96)
 
\bibitem[\protect\citeauthoryear{Haehnelt et al.}{2001}]{b1} Haehnelt
  M.G., Madau P., Kudritzki R., Haardt F., 2001, ApJ, 549, L151
   
 
\bibitem[\protect\citeauthoryear{Hui et al.}{2002}]{b83} Hui L.,
  Haiman Z., Zaldarriaga M., Alexander T., 2002, ApJ, 564, 525
  
\bibitem[\protect\citeauthoryear{Hui \& Gnedin}{1997}]{b2} Hui L.,
  Gnedin N., 1997, MNRAS, 292, 27
  
\bibitem[\protect\citeauthoryear{Hunt et al.}{2004}]{b89} Hunt M.P.,
  Steidel C.C., Adelberger K.L., Shapley A.E., 2004, ApJ, 605, 625


    
\bibitem[\protect\citeauthoryear{Hernquist et al.}{1996}]{b56}
  Hernquist L., Katz N., Weinberg D.H., Miralda-Escud\'{e} J., 1996,
  ApJ, 457, L51
  
\bibitem[\protect\citeauthoryear{Katz, Weinberg \&
    Hernquist}{1996}]{b11} Katz N., Weinberg D.H., Hernquist L., 1996,
  ApJS, 105, 19
  
\bibitem[\protect\citeauthoryear{Kim, Cristiani \&
    D'Odorico}{2001}]{b45} Kim T.-S., Cristiani S., D'Odorico S.,
  2001, A\&A, 373, 757
  
\bibitem[\protect\citeauthoryear{Kim et al.}{2002}]{b15} Kim T.-S.,
  Carswell R.F., Cristiani S., D'Odorico S., Giallongo E., 2002,
  MNRAS, 335, 555
  
\bibitem[\protect\citeauthoryear{Kirkman \& Tytler}{1997}]{b59}
  Kirkman D., Tytler D., 1997, ApJ, 484, 672
  
\bibitem[\protect\citeauthoryear{Leitherer et al.}{1995}]{b66}
  Leitherer C., Ferguson H.C., Heckman T.M., Lowenthal J.D., 1995,
  ApJ, 454, L19
  
\bibitem[\protect\citeauthoryear{Lu et al.}{1996}]{b14} Lu L., Sargent
  W.L.W., Womble D.S., Takada-Hidai M., 1996, ApJ, 472, 509
  
\bibitem[\protect\citeauthoryear{Madau, Haardt \& Rees}{1999}]{b37}
  Madau P., Haardt F., Rees M.J., 1999, ApJ, 585, 34 (MHR99)
  
\bibitem[\protect\citeauthoryear{McDonald et al.}{2001}]{b48} McDonald
  P., Miralda-Escud\'{e} J., Rauch M., Sargent W.L.W., Barlow T.A.,
  Cen R., 2001, ApJ, 562, 52
  
\bibitem[\protect\citeauthoryear{McDonald \&
    Miralda-Escud\'{e}}{2001}]{b18} McDonald P., Miralda-Escud\'{e}
  J., 2001, ApJ, 549, L11
  
\bibitem[\protect\citeauthoryear{Meiksin \& White}{2001}]{b60} Meiksin
  A., White M., 2001, MNRAS, 324, 141
  
\bibitem[\protect\citeauthoryear{Meiksin \& White}{2003}]{b54} Meiksin
  A., White M., 2003, MNRAS, 342, 1205
  
\bibitem[\protect\citeauthoryear{Meiksin \& White}{2004}]{b55} Meiksin
  A., White M., 2004, MNRAS, 350, 1107
  
\bibitem[\protect\citeauthoryear{Miralda Escud\'{e} et
    al.}{1996}]{b58} Miralda Escud\'{e} J., Cen R., Ostriker J.P.,
  Rauch M., 1996, ApJ, 471, 582
  
\bibitem[\protect\citeauthoryear{Norman \& Bryan}{1999}]{b44} Norman
  M.L., Bryan G.L., 1999, in: ASSL Vol. 240: Numerical Astrophysics,
  19
  
\bibitem[\protect\citeauthoryear{O'Shea et al.}{2003}]{b68} O'Shea
  B.W., Nagamine K., Springel V., Hernquist L., Norman M.L., 2003,
  preprint (astro-ph/0301265)
  
\bibitem[\protect\citeauthoryear{Press, Rybicki \&
    Schneider}{1993}]{b40} Press W.H., Ribicki G.B., Schneider D.P.,
  1993, ApJ, 414, 64
  
\bibitem[\protect\citeauthoryear{Rauch et al.}{1997}]{b3} Rauch M. et
  al., 1997, ApJ, 489, 7
  
\bibitem[\protect\citeauthoryear{Rauch}{1998}]{b97} Rauch M., 1998,
  ARA\&A, 36, 267
  
\bibitem[\protect\citeauthoryear{Reimers et al.}{1997}]{b46} Reimers
  D., Kohler S., Wisotzki L., Groote D., Rodriguez-Pascual P.,
  Wamsteker W., 1997, A\&A, 327, 890
  
\bibitem[\protect\citeauthoryear{Ricotti et al.}{2000}]{b29} Ricotti
  M., Gnedin N., Shull M., 2000, ApJ, 534, 41
   
\bibitem[\protect\citeauthoryear{Scott et al.}{2000}]{b24} Scott J.,
  Bechtold J., Dobrzycki A., Kulkarni V.P., 2000, ApJS, 130, 67
  
\bibitem[\protect\citeauthoryear{Scott et al.}{2004}]{b84} Scott J.E.,
  Kriss, G.A., Brotherton M., Green R.F., Hutchings J., Shull J.M.,
  Zheng W., 2004, ApJ, 615, 135
  
\bibitem[\protect\citeauthoryear{Schaye et al.}{2000}]{b27} Schaye J.,
  Theuns T., Rauch M., Efstathiou G., Sargent W.L.W., 2000, MNRAS,
  318, 817
  
\bibitem[\protect\citeauthoryear{Schaye et al.}{2003}]{b41} Schaye J.,
  Aguirre A., Kim T.-S., Theuns T., Rauch M., Sargent W.L.W., 2003,
  ApJ, 596, 768
  
\bibitem[\protect\citeauthoryear{Sokasian, Abel \&
    Hernquist}{2003}]{b86} Sokasian A., Abel T., Hernquist L., 2003,
  MNRAS, 340, 473
 
\bibitem[\protect\citeauthoryear{Songalia \& Cowie}{2002}]{b49}
  Songalia A., Cowie L.L., 2002, AJ, 123, 2183
  
\bibitem[\protect\citeauthoryear{Spergel et al.}{2003}]{b80} Spergel
  et al., 2003, ApJS, 148, 175
 
\bibitem[\protect\citeauthoryear{Springel \& Hernquist}{2002}]{b26}
  Springel V., Hernquist L., 2002, MNRAS, 333, 649
  
\bibitem[\protect\citeauthoryear{Springel \& Hernquist}{2003}]{b32}
  Springel V., Hernquist L., 2003, MNRAS, 339, 289
  
\bibitem[\protect\citeauthoryear{Springel, Yoshida \&
    White}{2001}]{b25} Springel V., Yoshida N., White S.D.M., 2001,
  NewA, 6, 79
  
\bibitem[\protect\citeauthoryear{Steidel et al.}{2001}]{b23} Steidel
  C.C., Pettini M., Adelberger K.L., 2001, ApJ, 546, 665
  
\bibitem[\protect\citeauthoryear{Theuns et al.}{1998}]{b12} Theuns T.,
  Leonard A., Efstathiou G., Pearce F., Thomas P.A., 1998, MNRAS, 301,
  478
  
\bibitem[\protect\citeauthoryear{Theuns et al.}{2000}]{b36} Theuns T.,
  Schaye J., Haehnelt M.G., 2000, MNRAS, 315, 600
  
\bibitem[\protect\citeauthoryear{Theuns et al.}{2002}]{b34} Theuns T.,
  Viel M., Kay S., Schaye J., Carswell B., Tzanavaris P., 2002, ApJ,
  578, L5
  
\bibitem[\protect\citeauthoryear{Tytler et al.}{2004}]{b5} Tytler D.
  et al., 2004, preprint (astro-ph/0403688)
    
\bibitem[\protect\citeauthoryear{Weinberg et al.}{1997}]{b52} Weinberg
  D.H., Hernquist L., Katz N., Croft R., Miralda-Escud\'{e} J., 1997,
  in P Petitjean \& S.  Charlot, eds, Proceedings of the 13th IAP
  Colloquium, Structure and Evolution of the Intergalactic Medium,
  p.133
  
\bibitem[\protect\citeauthoryear{Weinberg et al.}{1999}]{b6} Weinberg
  D. et al., 1999, in A.J. Banday, R.K. Sheth, L.N. da Costa, eds,
  Evolution of large scale structure: from recombination to Garching,
  p.346
  
\bibitem[\protect\citeauthoryear{Viel et al.}{2004}]{b69} Viel M.,
  Haehnelt M.G., Carswell R.F., Kim T.-S., 2004, MNRAS, 349, L33
  
\bibitem[\protect\citeauthoryear{Viel, Haehnelt \&
    Springel}{2004}]{b8} Viel M., Haehnelt M.G., Springel V., 2004, MNRAS, 354, 184
  
\bibitem[\protect\citeauthoryear{Zhang, Anninos \& Norman}{1995}]{b57}
  Zhang Y., Anninos P., Norman M.L., 1995, ApJ, 453, L57
  
\bibitem[\protect\citeauthoryear{Zhang et al.}{1998}]{b10} Zhang Y.,
  Meiksin A., Anninos P., Norman M.L., 1998, ApJ, 495, 63

\end{thebibliography}
\end{document}